\input harvmac
\input epsf
\input amssym
%\draftmode
%
\noblackbox
%%%%%%%%%%%%%%%%%%%%%%%%%%%%%%%%%%%%%%%%%%%%%%
%%%%%%%%%%%%%%%%%%%%%%%%%%%
% some stuff needed for figures:
%%%%%%%%%%%%%%%%%%%%%%%%%%%%%%%%%%%%%%%%%%%%%%
%%%%%%%%%%%%%%%%%%%%%%%%%%%
\newcount\figno
\figno=0
\def\fig#1#2#3{
\par\begingroup\parindent=0pt\leftskip=1cm\rightskip=1cm\parindent=0pt
\baselineskip=11pt \global\advance\figno by 1 \midinsert
\epsfxsize=#3 \centerline{\epsfbox{#2}} \vskip -21pt {\bf Fig.\
\the\figno: } #1\par
\endinsert\endgroup\par
}
\def\figlabel#1{\xdef#1{\the\figno}}
\def\encadremath#1{\vbox{\hrule\hbox{\vrule\kern8pt\vbox{\kern8pt
\hbox{$\displaystyle #1$}\kern8pt} \kern8pt\vrule}\hrule}}
%%%%%%%%%%%%%%%%%%%%%%%%%%%%%%%%%%%%%%%%%%%%%%
%%%%%%%%%%%%%%%%%%%%%%%%%%%
% definitions
%%%%%%%%%%%%%%%%%%%%%%%%%%%%%%%%%%%%%%%%%%%%%%
%%%%%%%%%%%%%%%%%%%%%%%%%%%

\def\frac#1#2{{#1 \over #2}}

\def\semi{\subset\kern-1em\times\;}
\def\bar#1{\overline{#1}}
\def\sqr#1#2{{\vcenter{\vbox{\hrule height.#2pt
\hbox{\vrule width.#2pt height#1pt \kern#1pt \vrule width.#2pt}
\hrule height.#2pt}}}}

 \def\cN{{\cal N}}    \def\cO{{\cal O}}

                   \def\cL{{\cal L}}
\def\cM{{\cal M}}                   \def\cN{{\cal N}}
\def\cO{{\cal O}}                   \def\cP{{\cal P}}
\def\cR{{\cal R}}                   
\def\cV{{\cal V}}                   \def\cW{{\cal W}}
\def\cZ{{\cal Z}}                   
\def\Neql#1{${\cal N} \! = \! #1$}
%
%%%%%%%%%%%%%%%%%%%%%%%%%%%%%%%%%%%%%%%%%%%%%%

%\def\oneone{\rlap 1\mkern4mu{\rm l}}
\def\coeff#1#2{\relax{\textstyle {#1 \over #2}}\displaystyle}
\def\ZZ{\Bbb{Z}}

\def\IR{\Bbb{R}}

%%%%%%%%%%%%%%%%%%%%%%%%%%%%%%%%%%%%%%%%%%%%%%
%%%%%%%%%%%%%%%%%%%%%%%%%%%%%%%%%%%%%%%%%%%%%%
% Definitions from Nick H

\def\al{\alpha}
\def\eps{\epsilon}
\def\tha{\theta}

\def\Gam{\Gamma}
\def\blp{{\big (}}
\def\brp{{\big )}}
\def\Blp{{\Big (}}
\def\Brp{{\Big )}}
\def\Dbar{{\bar D}}
\def\cZbar{{\bar \cZ}}
\def\tcZ{{\tilde \cZ}}
\def\tcZbar{{\overline \tcZ}}

%%%%%%%%%%%%%%%%%%%%%%%%%%%%%%%%%%%%%%%%%%%%%%
%%%  Krzysztof's definitions
%%%%%%%%%%%%%%%%%%%%%%%%%%%%%%%%%%%%%%%%%%%%%%
\ifx\pdfoutput\undefined

\else

\fi
%%%%%%%%%%%%%%%%%%%%%%%%%%%%%%%%%%%%%%%%%%%%%%

% Continuous colors

% Gothic fonts for Lie algebras

\font\teneufm=eufm10 \font\seveneufm=eufm7 \font\fiveeufm=eufm5
\newfam\eufmfam
\textfont\eufmfam=\teneufm \scriptfont\eufmfam=\seveneufm
\scriptscriptfont\eufmfam=\fiveeufm

\def\eps{{\epsilon}}

%%%%%%%%%%%%%%%%%%%%%%%%%%%%%%%%%%%%%%%%%%%%%%
%%%  Nikolay's definitions
%%%%%%%%%%%%%%%%%%%%%%%%%%%%%%%%%%%%%%%%%%%%%%
%%%%%%%%%%%%%%%%%%%%%%%%%%%

%%%%%%%%%%%%%%%%%%%%%%%%%%%%

%%%%%%%%%%%%%%%%%%%%%%%%%%%%%%%%%%%%%%%%%%%%%%
%%%%%%%%%%%%%%%%%%%%%%%%%%%
% References
%%%%%%%%%%%%%%%%%%%%%%%%%%%%%%%%%%%%%%%%%%%%%%
%%%%%%%%%%%%%%%%%%%%%%%%%%%
%

%\SeibergAX
\lref\SeibergAX{
  N.~Seiberg,
``Notes on theories with 16 supercharges,''
  Nucl.\ Phys.\ Proc.\ Suppl.\  {\bf 67}, 158 (1998)
  [arXiv:hep-th/9705117].
  %%CITATION = NUPHZ,67,158;%%
}

%\CorradoNV
\lref\CorradoNV{
  R.~Corrado, K.~Pilch and N.~P.~Warner,
``An N = 2 supersymmetric membrane flow,''
  Nucl.\ Phys.\  B {\bf 629}, 74 (2002)
  [arXiv:hep-th/0107220].
  %%CITATION = NUPHA,B629,74;%%
}

%\SchwarzYJ
\lref\SchwarzYJ{
  J.~H.~Schwarz,
  ``Superconformal Chern-Simons theories,''
  JHEP {\bf 0411}, 078 (2004)
  [arXiv:hep-th/0411077].
  %%CITATION = JHEPA,0411,078;%%
}

%\BaggerSK
\lref\BaggerSK{
  J.~Bagger and N.~Lambert,
  ``Modeling multiple M2's,''
  Phys.\ Rev.\  D {\bf 75}, 045020 (2007)
  [arXiv:hep-th/0611108].
  %%CITATION = PHRVA,D75,045020;%%
}

%\GustavssonVU
\lref\GustavssonVU{
  A.~Gustavsson,
  ``Algebraic structures on parallel M2-branes,''
  arXiv:0709.1260 [hep-th].
  %%CITATION = ARXIV:0709.1260;%%
}

%\BaggerJR
\lref\BaggerJR{
  J.~Bagger and N.~Lambert,
  ``Gauge Symmetry and Supersymmetry of Multiple M2-Branes,''
  Phys.\ Rev.\  D {\bf 77}, 065008 (2008)
  [arXiv:0711.0955 [hep-th]].
  %%CITATION = PHRVA,D77,065008;%%
}

%\BaggerVI
\lref\BaggerVI{
  J.~Bagger and N.~Lambert,
  ``Comments On Multiple M2-branes,''
  JHEP {\bf 0802}, 105 (2008)
  [arXiv:0712.3738 [hep-th]].
  %%CITATION = JHEPA,0802,105;%%
}

%\AharonyUG
\lref\AharonyUG{
  O.~Aharony, O.~Bergman, D.~L.~Jafferis and J.~Maldacena,
  ``N=6 superconformal Chern-Simons-matter theories, M2-branes and their gravity duals,''
  JHEP {\bf 0810}, 091 (2008)
  [arXiv:0806.1218 [hep-th]].
  %%CITATION = JHEPA,0810,091;%%
}
%
%\fAI
\lref\MauriAI{
  A.~Mauri and A.~C.~Petkou,
 ``An N=1 Superfield Action for M2 branes,''
  Phys.\ Lett.\  B {\bf 666}, 527 (2008)
  [arXiv:0806.2270 [hep-th]].
  %%CITATION = PHLTA,B666,527;%%
}
%
%\BennaZY
\lref\BennaZY{
  M.~Benna, I.~Klebanov, T.~Klose and M.~Smedback,
  ``Superconformal Chern-Simons Theories and $AdS_4/CFT_3$ Correspondence,''
  JHEP {\bf 0809}, 072 (2008)
  [arXiv:0806.1519 [hep-th]].
  %%CITATION = JHEPA,0809,072;%%
}
%
%\AvdeevZA
\lref\AvdeevZA{
  L.~V.~Avdeev, G.~V.~Grigorev and D.~I.~Kazakov,
  ``Renormalizations in Abelian Chern-Simons field theories with matter,''
  Nucl.\ Phys.\  B {\bf 382}, 561 (1992).
  %%CITATION = NUPHA,B382,561;%%
}
%
%\AvdeevJT
\lref\AvdeevJT{
  L.~V.~Avdeev, D.~I.~Kazakov and I.~N.~Kondrashuk,
  ``Renormalizations in supersymmetric and nonsupersymmetric nonAbelian
  Chern-Simons field theories with matter,''
  Nucl.\ Phys.\  B {\bf 391}, 333 (1993).
  %%CITATION = NUPHA,B391,333;%%
}
%
%\KlebanovVQ
\lref\KlebanovVQ{
  I.~Klebanov, T.~Klose and A.~Murugan,
  ``$AdS_4/CFT_3$ -- Squashed, Stretched and Warped,''
  arXiv:0809.3773 [hep-th].
  %%CITATION = ARXIV:0809.3773;%%
}
%
%\VanRaamsdonkFT
\lref\VanRaamsdonkFT{
  M.~Van Raamsdonk,
 ``Comments on the Bagger-Lambert theory and multiple M2-branes,''
  JHEP {\bf 0805}, 105 (2008)
  [arXiv:0803.3803 [hep-th]].
  %%CITATION = JHEPA,0805,105;%%
}

%\deWitIG
\lref\deWitIG{
  B.~de Wit and H.~Nicolai,
``N=8 Supergravity,''
  Nucl.\ Phys.\  B {\bf 208}, 323 (1982).
  %%CITATION = NUPHA,B208,323;%%
}
%
%\WarnerVZ
\lref\WarnerVZ{
  N.~P.~Warner,
``Some New Extrema Of The Scalar Potential Of Gauged N=8
Supergravity,''
  Phys.\ Lett.\  B {\bf 128}, 169 (1983).
  %%CITATION = PHLTA,B128,169;%%
}
\lref\WarnerDU{
  N.~P.~Warner,
``Some Properties Of The Scalar Potential In Gauged Supergravity
Theories,''
  Nucl.\ Phys.\  B {\bf 231}, 250 (1984).
  %%CITATION = NUPHA,B231,250;%%
}
%
 %\AhnMF
\lref\AhnMF{
  C.~h.~Ahn and K.~Woo,
  ``Supersymmetric domain wall and RG flow from 4-dimensional gauged N = 8 supergravity,''
  Nucl.\ Phys.\  B {\bf 599}, 83 (2001)
  [arXiv:hep-th/0011121].
  %%CITATION = NUPHA,B599,83;%%
}
%
%\AhnBY
\lref\AhnBY{
  C.~h.~Ahn and K.~s.~Woo,
``Domain wall and membrane flow from other gauged d = 4, N = 8
supergravity.
  %I,''
  Nucl.\ Phys.\  B {\bf 634}, 141 (2002)
  [arXiv:hep-th/0109010].
  %%CITATION = NUPHA,B634,141;%%
}
%
%\AhnQGA
\lref\AhnQGA{
  C.~h.~Ahn and K.~s.~Woo,
  ``Domain wall from gauged d = 4, N = 8 supergravity. II,''
  JHEP {\bf 0311}, 014 (2003)
  [arXiv:hep-th/0209128].
  %%CITATION = JHEPA,0311,014;%%
}
%\AhnAQ
\lref\AhnAQ{
  C.~h.~Ahn and J.~Paeng,
``Three-dimensional SCFTs, supersymmetric domain wall and
renormalization group flow,''
  Nucl.\ Phys.\  B {\bf 595}, 119 (2001)
  [arXiv:hep-th/0008065].
  %%CITATION = NUPHA,B595,119;%%
}
%
%\AhnKW
\lref\AhnKW{
  C.~h.~Ahn and T.~Itoh,
``An N = 1 supersymmetric G(2)-invariant flow in M-theory,''
  Nucl.\ Phys.\  B {\bf 627}, 45 (2002)
  [arXiv:hep-th/0112010].
  %%CITATION = NUPHA,B627,45;%%
}
%
%\AhnQGA
\lref\AhnQGA{
  C.~h.~Ahn and K.~s.~Woo,
 ``Domain wall from gauged d = 4, N = 8 supergravity. II,''
  JHEP {\bf 0311}, 014 (2003)
  [arXiv:hep-th/0209128].
  %%CITATION = JHEPA,0311,014;%%
}

%\AhnYA
\lref\AhnYA{
  C.~Ahn,
  ``Holographic Supergravity Dual to Three Dimensional N=2 Gauge Theory,''
  JHEP {\bf 0808}, 083 (2008)
  [arXiv:0806.1420 [hep-th]].
  %%CITATION = JHEPA,0808,083;%%
}

%\AhnGDA
\lref\AhnGDA{
  C.~Ahn,
  ``Towards Holographic Gravity Dual of N=1Superconformal Chern-Simons Gauge
  Theory,''
  JHEP {\bf 0807}, 101 (2008)
  [arXiv:0806.4807 [hep-th]].
  %%CITATION = JHEPA,0807,101;%%
}

%\AhnJJ
\lref\AhnJJ{
  C.~Ahn,
  ``Comments on Holographic Gravity Dual of N=6 Superconformal Chern-Simons
  Gauge Theory,''
  arXiv:0812.4363 [hep-th].
  %%CITATION = ARXIV:0812.4363;%%
}

%\FreedmanGP
\lref\FreedmanGP{
  D.~Z.~Freedman, S.~S.~Gubser, K.~Pilch and N.~P.~Warner,
``Renormalization group flows from holography supersymmetry and a
c-theorem,''
  Adv.\ Theor.\ Math.\ Phys.\  {\bf 3}, 363 (1999)
  [arXiv:hep-th/9904017].
  %%CITATION = 00203,3,363;%%
}
%

%\WarnerKH
\lref\WarnerKH{
  N.~P.~Warner,
``Renormalization group flows from five-dimensional
supergravity,''
  Class.\ Quant.\ Grav.\  {\bf 17}, 1287 (2000)
  [arXiv:hep-th/9911240].
  %%CITATION = CQGRD,17,1287;%%
}
%
%\KhavaevGB
\lref\KhavaevGB{
  A.~Khavaev and N.~P.~Warner,
``A class of N = 1 supersymmetric RG flows from five-dimensional N
= 8 supergravity,''
  Phys.\ Lett.\  B {\bf 495}, 215 (2000)
  [arXiv:hep-th/0009159].
  %%CITATION = PHLTA,B495,215;%%
}
%
 %\FreedmanGK
\lref\FreedmanGK{
  D.~Z.~Freedman, S.~S.~Gubser, K.~Pilch and N.~P.~Warner,
``Continuous distributions of D3-branes and gauged supergravity,''
  JHEP {\bf 0007}, 038 (2000)
  [arXiv:hep-th/9906194].
  %%CITATION = JHEPA,0007,038;%%
}
%

%\NicolaiHS
\lref\NicolaiHS{
  H.~Nicolai and N.~P.~Warner,
  ``The SU(3) X U(1) Invariant Breaking Of Gauged N=8 Supergravity,''
  Nucl.\ Phys.\  B {\bf 259}, 412 (1985).
  %%CITATION = NUPHA,B259,412;%%
}

\lref\BenaZB{
  I.~Bena,
  ``The M-theory dual of a 3 dimensional theory with reduced supersymmetry,''
  Phys.\ Rev.\  D {\bf 62}, 126006 (2000)
  [arXiv:hep-th/0004142].
  %%CITATION = PHRVA,D62,126006;%%
}

%
%\LinNB
\lref\LinNB{
  H.~Lin, O.~Lunin and J.~M.~Maldacena,
``Bubbling AdS space and 1/2 BPS geometries,''
  JHEP {\bf 0410}, 025 (2004)
  [arXiv:hep-th/0409174].
  %%CITATION = JHEPA,0410,025;%%
}

\lref\GomisVC{
  J.~Gomis, D.~Rodriguez-Gomez, M.~Van Raamsdonk and H.~Verlinde,
  ``A Massive Study of M2-brane Proposals,''
  JHEP {\bf 0809}, 113 (2008)
  [arXiv:0807.1074 [hep-th]].
  %%CITATION = JHEPA,0809,113;%%
}
%
%\PopeJP
\lref\PopeJP{
  C.~N.~Pope and N.~P.~Warner,
``A dielectric flow solution with maximal supersymmetry,''
  JHEP {\bf 0404}, 011 (2004)
  [arXiv:hep-th/0304132].
  %%CITATION = JHEPA,0404,011;%%
}
\lref\BenaJW{
  I.~Bena and N.~P.~Warner,
  ``A harmonic family of dielectric flow solutions with maximal
  supersymmetry,''
  JHEP {\bf 0412}, 021 (2004)
  [arXiv:hep-th/0406145].
  %%CITATION = JHEPA,0412,021;%%
}
\lref\GomisCV{
  J.~Gomis, A.~J.~Salim and F.~Passerini,
  ``Matrix Theory of Type IIB Plane Wave from Membranes,''
  JHEP {\bf 0808}, 002 (2008)
  [arXiv:0804.2186 [hep-th]].
  %%CITATION = JHEPA,0808,002;%%
}
%
%\PilchUE
\lref\PilchUE{
  K.~Pilch and N.~P.~Warner,
  ``N = 2 supersymmetric RG flows and the IIB dilaton,''
  Nucl.\ Phys.\  B {\bf 594}, 209 (2001)
  [arXiv:hep-th/0004063].
  %%CITATION = NUPHA,B594,209;%%
}

%\GowdigereUK
\lref\GowdigereUK{
  C.~N.~Gowdigere and N.~P.~Warner,
``Flowing with eight supersymmetries in M-theory and F-theory,''
  JHEP {\bf 0312}, 048 (2003)
  [arXiv:hep-th/0212190].
  %%CITATION = JHEPA,0312,048;%%
}

%\GowdigereJF
\lref\GowdigereJF{
  C.~N.~Gowdigere, D.~Nemeschansky and N.~P.~Warner,
``Supersymmetric solutions with fluxes from algebraic Killing spinors,''
  Adv.\ Theor.\ Math.\ Phys.\  {\bf 7}, 787 (2004)
  [arXiv:hep-th/0306097].
  %%CITATION = 00203,7,787;%%
}

%\DonagiCF
\lref\DonagiCF{
  R.~Donagi and E.~Witten,
  ``Supersymmetric Yang-Mills Theory And Integrable Systems,''
  Nucl.\ Phys.\  B {\bf 460}, 299 (1996)
  [arXiv:hep-th/9510101].
  %%CITATION = NUPHA,B460,299;%%
}

%\deWitNZ
\lref\deWitNZ{
  B.~de Wit, H.~Nicolai and N.~P.~Warner,
  ``The Embedding Of Gauged N=8 Supergravity Into D = 11 Supergravity,''
  Nucl.\ Phys.\  B {\bf 255}, 29 (1985).
  %%CITATION = NUPHA,B255,29;%%
}

%\DoreySJ
\lref\DoreySJ{
  N.~Dorey,
``An elliptic superpotential for softly broken N = 4 supersymmetric Yang-Mills theory,''
  JHEP {\bf 9907}, 021 (1999)
  [arXiv:hep-th/9906011].
  %%CITATION = JHEPA,9907,021;%%
}

%\DoreyFC
\lref\DoreyFC{
  N.~Dorey and S.~P.~Kumar,
``Softly-broken N = 4 supersymmetry in the large-N limit,''
  JHEP {\bf 0002}, 006 (2000)
  [arXiv:hep-th/0001103].
  %%CITATION = JHEPA,0002,006;%%
}

%\deWitIY
\lref\deWitIY{
  B.~de Wit and H.~Nicolai,
``The Consistency of the S**7 Truncation in D=11 Supergravity,''
  Nucl.\ Phys.\  B {\bf 281}, 211 (1987).
  %%CITATION = NUPHA,B281,211;%%
}

%
%%%%%%%%%%%%% End References %%%%%%%%%%%%%
%%%%%%%%%%%%%%%%%%%%%%%%%%%%%%%%%%%%%%%%

%%%%%%%%%%%%%%%   Title Page  %%%%%%%%%%%%%
 
\Title{ \vbox{ \hbox{NSF-KITP-09-05 }}} { \vbox{\vskip -1.5cm
\centerline{\hbox{Holographic, ${\cN \!=\! 1}$ Supersymmetric RG
Flows}} \vskip 0.2cm \centerline{\hbox{on M2 Branes }}}} \vskip
-0.9cm \centerline{Nikolay~Bobev${}^{(1)}{}^{(2)}$,  Nick
Halmagyi${}^{(3)}$${}^{(4)}$${}^{(5)}$, Krzysztof Pilch${}^{(1)}$
and Nicholas P.\ Warner${}^{(1)}$}

\bigskip
\centerline{{${}^{(1)}$\it Department of Physics and Astronomy,
University of Southern California}} \centerline{{\it Los Angeles,
CA 90089-0484, USA}}
\medskip
\centerline{{${}^{(2)}$\it Kavli Institute for Theoretical
Physics, University of California}}\centerline{{\it Santa Barbara
CA 93106-4030, USA}}
\medskip
\centerline{{${}^{(3)}$\it Laboratoire de Physique Th\' eorique et
Hautes Energies,  CNRS UMR 7589}} \centerline{{\it Universit\' e
Pierre et Marie Curie, 4 Place Jussieu, 75252 Paris Cedex 05,
France }}
\medskip \centerline{{${}^{(4)}$\it Laboratoire de
Physique Th\' eorique de l'\' Ecole Normale Sup\' erieure,}}
\centerline {{\it24 rue Lhomond, 75231 Paris, France }}
\medskip
\centerline{{${}^{(5)}$\it Institut de Physique Th\' eorique,
CEA/Saclay, CNRS-URA 2306,}} \centerline{{\it Orme des Merisiers,
F-91191 Gif sur Yvette, France}}
\bigskip
\bigskip
We find a family of holographic  ${\cN \!=\! 1}$ supersymmetric RG
flows on M2 branes.  These flows are driven by two mass parameters
from the maximally (${\cN \!=\! 8}$) supersymmetric theory and the
infra-red theory is controlled by two fixed points, one with $G_2$
symmetry and the other with  $SU(3) \times U(1)$ symmetry and
${\cN \!=\! 2}$ supersymmetry.  The generic flow, with unequal
mass parameters, is ${\cN \!=\! 1}$ supersymmetric but goes to the
$SU(3) \times U(1)$ symmetric, ${\cN \!=\! 2}$ supersymmetric
fixed point, where the masses are equal.    The only flow that
goes to the $G_2$ symmetric point occurs when one of the mass
parameters is set to zero.   There is an ${\cN \!=\! 1}$
supersymmetric flow from the $G_2$ symmetric point to the $SU(3)
\times U(1)$ symmetric point and supergravity gives a prediction
of $\pm {1 \over \sqrt{6}}$ for the anomalous dimensions of the
operators that drive this flow.  We examine these flows from the
field theory perspective but find that one is limited to
qualitative results since ${\cN \!=\! 1}$ supersymmetry in three
dimensions is insufficient to protect the form and dimensions of
the operators involved  in the flow.

%\draft
\Date{\sl {January, 2009}}

%\vfill\eject

%%%%%%%%%%%%%%%%%%%%%%%%%%%%%%%%%%
\newsec{Introduction}
%%%%%%%%%%%%%%%%%%%%%%%%%%%%%%%%%%

The field theory on M2 branes has always posed something of a
problem in that in its simplest, most supersymmetric formulation,
it is necessarily strongly coupled \SeibergAX\ .  As a consequence,
some of the non-trivial results about infra-red flows and fixed
points in this theory were first obtained via holography.  It was known
from much older work on four-dimensional gauged supergravity
\WarnerVZ\ that the maximally supersymmetric (\Neql8) theory must
have two non-trivial supersymmetric infra-red fixed points
corresponding to massive flows in the field theory. Some of these
holographic flows were explicitly constructed in gauged
supergravity \refs{\AhnMF\AhnBY\AhnQGA\AhnAQ{--}\AhnKW} and
directly in M-theory \CorradoNV\ (see also
\refs{\AhnYA\AhnGDA{--}\AhnJJ} for more recent work).  More
generally, there are quite a number of known supersymmetric
compactifications of M-theory that correspond to M2 brane
configurations and might therefore be incorporated into a web of
RG flows that connect to the \Neql8 theory. While there were some
interesting parallels between the M2-brane theory and the
compactification of \Neql4 Yang-Mills theory, progress in this
area was limited by the lack of a good field theory description on
the M2 brane.

Recently our understanding of the underlying M2-brane field theory
has vastly improved
\refs{\SchwarzYJ\BaggerSK\GustavssonVU\BaggerJR\BaggerVI\VanRaamsdonkFT{--}\AharonyUG}.
This theory may be understood in terms of an \Neql6
Chern-Simons-matter theory in which the  level, $k$, emerges from
a $\ZZ_k$ orbifold. That is, if one takes the compactifying
manifold to be $S^7/\ZZ_k$, where the $\ZZ_k$ acts on the Hopf
fiber, then for $k>2$ the supersymmetry is broken to \Neql6 and
the $\cR$-symmetry is broken from $SO(8)$ to $SO(6)$.  The
coupling of this field theory may be thought of as $k^{-1}$, and
so it is weakly coupled for large $k$. For $k=1,2$ the full \Neql8
supersymmetry and $SO(8)$ $\cR$-symmetry is preserved (but not
manifestly within the ABJM formalism) but the theory is strongly
coupled.   This formulation has enabled one to re-examine and
understand the supergravity flows from the field-theory
perspective \refs{\BennaZY, \KlebanovVQ}.

In this paper we will exhibit and study a family of \Neql1
supersymmetric RG flows using the maximally supersymmetric \Neql8
theory.  This family is controlled by two infra-red fixed points:
\medskip
\item{\bf I}:  A fixed point with \Neql1 supersymmetry and a global  $G_2$ symmetry.   The flow corresponds to turning on a single mass parameter, with the remaining massless bosons and fermions on the M2 brane transforming in the ${\bf 7}$ of $G_2$.
\item{\bf II}:  A fixed point with  \Neql2 supersymmetry, a $U(1)$ $\cR$-symmetry and a global $SU(3)$  symmetry.   The flow corresponds to turning on two equal mass parameters, with the six remaining massless bosons and fermions  on the M2 brane transforming in the ${\bf 3} + {\bf \bar 3}$ of $SU(3)$.
\medskip
This family of flows is driven by two {\it independent} mass
parameters arranged so that when one of them vanishes the theory
flows in a $G_2$ invariant manner to fixed point {\bf I}  and when
they are equal  the theory flows in a $SU(3) \times U(1)$
invariant manner to fixed point {\bf II}. We will show that the
dominant fixed point is the ``lower''  fixed point, {\bf II}. That
is, the generic flow with two unequal masses flows to fixed point
{\bf II} where the two masses become equal and one only reaches
fixed point {\bf I}  if one of the mass parameters is exactly
zero. If one of the mass parameters is tiny compared to the other
then the flow can approach fixed point {\bf I} arbitrarily closely
before diverting to fixed point {\bf II}. There is also a flow
directly from fixed point {\bf I} to fixed point {\bf II}
directly, and the supergravity gives a prediction of $\pm {1 \over
\sqrt{6}}$ for the anomalous dimensions of the relevant operators
that drive this flow.

We also examine these flows using the field theory.    Indeed, the
\Neql2 supersymmetric flow and fixed point has been extensively
studied in \refs{\BennaZY, \KlebanovVQ} and the results have been
matched directly with supergravity \NicolaiHS. The field theory
description of the family of \Neql1 flows is rather more
qualitative.  This is because the \Neql2 flows are driven by
F-terms and are thus based upon operators whose dimensions and
interactions are protected by the $U(1)$ $\cR$-symmetry.   In the
\Neql1 flows, the action consists only of D-terms and these
generally undergo non-trivial renormalizations. We discuss the
usual procedure of integrating out the massive fields and find
that it gives some reasonable qualitative results that match the
supergravity, but the \Neql1 supersymmetry limits the analysis
significantly and does not show that generic mass perturbations
flow to  fixed point {\bf II}.  It also remains unclear how one
might compute the anomalous dimensions of $\pm {1 \over \sqrt{6}}$
in the field theory.  We thus have supergravity predictions that
pose an interesting challenge for field theory.

In Section 2 we  discuss the field theory underlying the family of
\Neql1 flows from the \Neql8 M2-brane theory and we present the
dual supergravity analysis in Section 3. We summarize our results
and conclude in Section 4. Some technical details of superfield
expansions have been put in the Appendix.

%%%%%%%%%%%%%%%%%%%%%%%%%%%%%%%%%%
\newsec{An $SU(3)$ invariant family of flows}
%%%%%%%%%%%%%%%%%%%%%%%%%%%%%%%%%%

We discuss a family of \Neql1  supersymmetric flows away from the
Bagger-Lambert-Gustavsson (BLG) theory, triggered by masses for
two real \Neql1 superfields. It is important to state the caveat
that by using the BLG theory, we are really studying two M2
branes: The gauge group of the BLG theory is $SU(2)\times SU(2)$
and so one is not exactly in the large-$N$ limit.    Nonetheless,
the intuition gathered from studying deformations of the BLG
action will prove to be confirmed in the gravity dual.   It is
also possible that some version of these dual supergravity
geometries will prove useful in more general ABJM theories.

It is also important to note that, for the class of supersymmetric
flows we study here, the supersymmetry will actually be completely
broken in the general ABJM model.  This is because the generic
ABJM theory has \Neql6 supersymmetry and an $SO(6)$ $\cR$-symmetry
and it is only for an $SU(2)\times SU(2)$ gauge group with $k=1,2$
that this symmetry is enhanced to \Neql8.  (The two extra
supersymmetries transform as $SO(6)$ singlets.)  The family of
flows we consider here break the \Neql8 supersymmetry to \Neql1
while preserving the $SU(3)$ subgroup of $SO(6)$.  This means that
our flows do not preserve any of the \Neql6 supersymmetries of the
ABJM theories and preserve one (or both) of the two $SO(6)$
singlet supersymmetries.  Thus we focus on the BLG theory for
which our supersymmetries are unbroken within the field theory.

%%%%%%%%%%%%%%%%%%%%%%
\subsec{The Bagger-Lambert-Gustavsson action in superspace}
%%%%%%%%%%%%%%%%%%%%%%

The BLG theory can be written in \Neql2 superspace with manifest
$SU(4)\times U(1)$ symmetry \refs{\BennaZY} and in \Neql1
superspace with $SO(7)$ symmetry \refs{\MauriAI}. One quirk of the
BLG formulation is that the gauge superfield $\cV_{ab}$ (which has
a component $A_{\mu ab}$) has to be contracted in two different
ways, and for this we define  ${\cV^a}_{b}$ and
${\hat{\cV}^a}{}_{b}$:
\eqn\VVh{ {\cV^{a}}_{b} ~=~ h^{ac}\cV_{cb}\,, \qquad
{\hat{\cV}}^{a}\,_{b}  ~=~  {f^{cda}}_b  \cV_{cd} \,. }
The tensor $h^{ac}$ is the non-degenerate bilinear form associated
with the three-alegbra and the components of the tensor
${f^{abc}}_d$ are the structure constants of the three-algebra
\refs{\BaggerJR}. The apparently unusual fact that we need both of
these contractions of the gauge field can be understood by
converting to an $SU(2)\times SU(2)$ gauge theory and re-writing
the theory in terms of bi-fundamental matter
\refs{\VanRaamsdonkFT}.   This description also requires the use
of complex combinations of the scalar superfields. The advantage
of formulating the BLG theory as a bi-fundamental gauge theory is
that it leads to the ABJM generalization, which  has gauge group
$U(N)\times U(N)$. However the gauge index structure of these
models, for $N>2$, does not allow for local holomorphic mass terms
whereas the original BLG theory does allow for such mass terms.

We therefore work with the BLG action\foot{We will always scale
fields such that the level appears as a factor multiplying the
whole action.}:
\eqn\SNTwo{ S_{BLG}= k(S_{CS} + S_{kin}+ W   ),}
where
\eqn\SNThree{\eqalign{ S_{CS}&= \int d^3x d^4 \tha \int_0^1 dt \,
\tr \Blp  \cV \Dbar^\al e^{it\hat{\cV}} D_\al e^{-it\hat{\cV}}
\Brp ,   \cr S_{kin}&=  \int d^3x d^4 \tha  \, \cZbar_A^a
e^{-2\hat{\cV}} \cZ^{A,a}, \cr W&=-\frac{1}{24} \eps_{ABCD}
\eps^{abcd} \int d^3x d^2\tha\, \cZ^A_a\cZ^B_b\cZ^C_c\cZ^D_d ~+~
c.c  \,. }}
We can now decompose the \Neql2 superspace expression above to
\Neql1  superspace. As described in the appendix, the $\tcZ^A$
fields are the $\cN=1$ projections of the $\cN=2$ superfields
$\cZ^A$. Using the complex structure on the \Neql1 superfields
\eqn\Zs{\eqalign{ \tcZ^1= \Phi^1+i \Phi ^2, &\ \ \  \tcZ^2= \Phi
^3+i \Phi ^4 \,, \cr \tcZ^3= \Phi ^5+i \Phi ^6, &\ \ \  \tcZ^4=
\Phi ^7+i \Phi ^8 \, }}
we recover the $SO(7)$ invariant description of BLG of
Mauri-Petkou \refs{\MauriAI},
\eqn\BLGCIJKL{\eqalign{ S_{BLG}&=\int d^3 x d^2 \tha \Blp -{1\over
2} \blp D^\al \Phi^I_b-{\eps^{abc}}_{d} \Gam^\al_{ab}  \Phi^I_c
\brp^2  -{1\over 8} \eps^{abcd} (D^\al \Gam^\beta_{ab} )(D_\beta
\Gam_{\al\, cd}) \brp \cr & -{1\over 6} {\eps^{cda}}_{g} \,
\eps^{efgh} (D^\al \Gam^\beta_{ab}) \Gam_{\al \, cd} \Gam_{\beta\,
ef} -\frac{1}{24} \eps^{abcd} C_{IJKL} \Phi^I_a \Phi^J_b \Phi^K_c
\Phi^L_d \Brp . }}
The tensor $C_{IJKL}$ is the self-dual $SO(7)$ invariant tensor
\eqn\CIJKL{\eqalign{ C_{IJKL} = & \Blp \delta^{1234}_{IJKL}  +
\delta^{5678}_{IJKL} + \delta^{1256}_{IJKL} + \delta^{3478}_{IJKL}
+ \delta^{3456}_{IJKL} + \delta^{1278}_{IJKL} \cr &
-(\delta^{1357}_{IJKL} + \delta^{2468}_{IJKL} )+
(\delta^{2457}_{IJKL} + \delta^{1368}_{IJKL}) +
(\delta^{1458}_{IJKL} +\delta^{2367}_{IJKL})  + (
\delta^{1467}_{IJKL} + \delta^{2358}_{IJKL}) \Brp \,. }}
Written this way we can identify the first line in \CIJKL\ as
coming from the \Neql2  $D$-terms in \SNTwo\ whereas the second
line in \CIJKL\ comes from the $F$-terms in \SNTwo.

Recall from the appendix that the \Neql2 vector multiplet $\cV$
contains an \Neql1 vector multiplet $\Gam^\al$ and an auxiliary
real \Neql1 scalar multiplet $R$. The $D$-term contributions to
the superpotential are obtained by integrating out $R$.

More precisely, we can write the \Neql1 superpotential as:
\eqn\WNone{\eqalign{
\widetilde{W}(\tcZ,\overline{\tcZ})&=-\frac{1}{24} \eps^{abcd}
C_{IJKL} \Phi^I_a \Phi^J_b \Phi^K_c \Phi^L_d \cr &=
\frac{1}{8}\eps^{abcd}\tcZbar_a^A \tcZ^A_b  \tcZbar_c^B \tcZ^B_d +
\frac{1}{48}\eps^{abcd}\eps_{ABCD}  \blp \tcZ^A_a \tcZ^B_b
\tcZ^C_c \tcZ^D_d  +\tcZbar^A_a \tcZbar^B_b \tcZbar^C_c
\tcZbar^D_d  \brp \,, }}
where, as described in the appendix, the $\tcZ^A$ fields are the
$\cN=1$ projections of the $\cN=2$ superfields $\cZ^A$ and the
$\Phi^I$ are the real components of the complex fields $\tcZ^A$.
The first (second) term in \WNone\ comes from the first (second)
line in \CIJKL\  and thus come from D(F)-terms of the \Neql2 action.

%%%%%%%%%%%%%%%%%%%%%%
\subsec{The family of RG flows}
%%%%%%%%%%%%%%%%%%%%%%

We now consider the deformation of the BLG action by adding mass
terms of the form:
\eqn\DWBLG{ \Delta W_{BLG} ~=~  \half m_7 \,  \Phi_7^2 ~+~  \half
m_8 \, \Phi_8^2\,.}
If we just give a mass to one of the fields ($m_7=0$ or $m_8=0$),
then we preserve $G_2$ symmetry. This is easily seen since the
bosons are in ${\bf 8_v}$ of $SO(8)$ and the spinors are in ${\bf
8_s}$, so giving a supersymmetric mass to one boson and one
fermion will preserve the subgroup $G_2$. If we give an unequal
mass to both fields, then the remaining symmetry is $SU(3)$ and if
we give an equal mass to both fields the symmetry is $SU(3)\times
U(1)$. With the extra $U(1)$ symmetry we preserve \Neql2
supersymmetry.

We find that the IR behavior of this family of flows is best
studied through the gravity dual and this is done in the next
section of this paper.  The picture that emerges is that if one of
the masses vanishes then the theory flows to a $G_2$ invariant
SCFT while for all other values of $(m_7,m_8)$ the theory flows to
the unique $SU(3)\times U(1)$ invariant point. In this sense the
$SU(3)\times U(1)$ point is a basin of attraction for these flows.

%%%%%%%%%%%%%%%%%%%%%%%%%%%%%%%
\goodbreak\midinsert
\vskip .2cm \centerline{ \epsfxsize 6cm \epsfbox{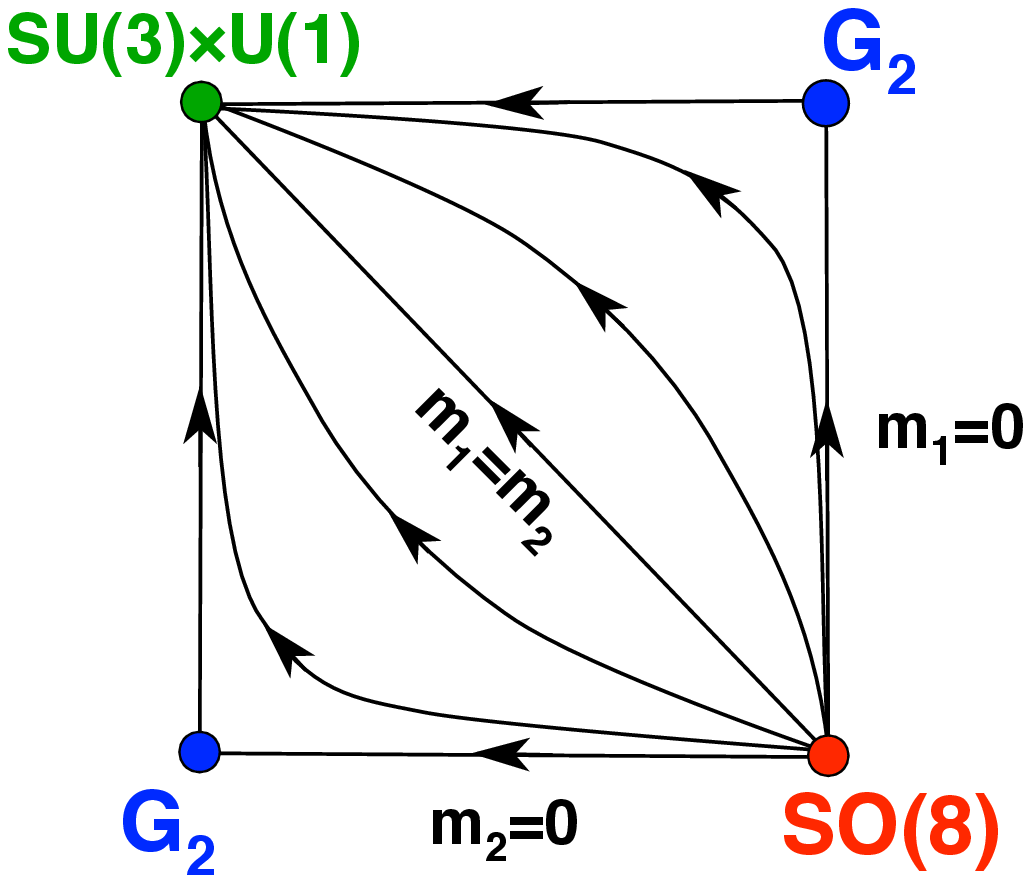} }
\vskip 0.2cm \leftskip 2pc
 \rightskip 2pc\noindent{\ninepoint\sl
{\bf Fig.~1}: \baselineskip= 8pt The pattern of RG flows to the
infra-red.  Starting from the $SO(8)$-invariant fixed point the
theory flows to a $G_2$-invariant fixed point only if one of the
masses vanishes.  If both masses are non-zero, but not necessarily
equal, then the theory flows to the $SU(3) \times U(1)$-invariant
fixed point where the masses become equal.  The two
$G_2$-invariant points are equivalent and there is a flow directly
from this fixed point  to the $SU(3) \times U(1)$-invariant fixed
point.}
\endinsert
%%%%%%%%%%%%%%%%%%%%%%%%%%%%%%%

Further, the gravity dual shows that there is a distinct RG flow
from the $G_2$ symmetric SCFT to the $SU(3)\times U(1)$ point
which preserves just $SU(3)$ along the flow. Given that such a
flow lies at the boundary of the family studied here, one is led
to conclude that this flow is triggered by the second mass term
however it is currently difficult to perform a mapping of the
operator spectrum in the field theory to the spectrum of
supergravity modes.  Indeed, the anomalous dimensions predicted by
the supergravity suggest that this operator mapping will be rather
non-trivial.

Whilst this detailed picture of the family of RG flows emerges
from the gravity dual, one can also get some intuition about the
family of flows by studying the field theory.  From the explicit
form of the \Neql1 superpotential \CIJKL\ one finds that, when
$m_7\ne m_8$,  the usual techniques of integrating out these
masses is problematic. For example, if we set $m_7=0$, then we can
seemingly integrate out $\Phi_8$  to obtain:
\eqn\WGTwo{ W_{G_2}\sim (\eps^{abcd}C_{IJK8}\Phi^I_a \Phi^J_b
\Phi^K_c)^2 +
 \eps^{abcd} \sum_{IJKL\neq8} C_{IJKL}\Phi^I_a \Phi^J_b \Phi^K_c \Phi^L_d \,,
}
which has terms quartic and sextic in the remaining seven fields.
Since these fields transform in the $\bf{7}$ of the unbroken
$G_2$, they must individually have equal dimension, and so one
might reasonably expect that the quartic and sextic terms to have
different dimensions and thus conclude that the resulting theory
cannot be conformal. This is not quite accurate since $\cN=1$
supersymmetry in three dimensions has no $\cR$-symmetry. As a
result one cannot conclude that the dimension of monomials in the
superpotential is simply the sum of the dimensions of each
component. We are thus left unable to determine the quantum
dimension of each term in \WGTwo\ since this is a strongly coupled
field theory.  For the \Neql8 theory one can take $k$ large and
study perturbation theory however as mentioned already, the flows
considered here are only supersymmetric for $k=1,2$.

On the other hand, if  $m_7=m_8$ one has  \Neql2 supersymmetry and
one can make a field-theory argument that the RG flow terminates
at a CFT fixed point in the IR \refs{\BennaZY}.  Indeed, when one
integrates out the $\cZ^4$ superfield one ends up with the
superpotential \refs{\BennaZY}:
\eqn\WNtwo{ W_{\cN=2}=\int d^3x d^2\tha (\eps_{abcd} \cZ_1^a
\cZ_2^b \cZ_3^c )^2 \,. }
At this point we have a $U(1)_R$ symmetry and thus we know that
the dimension of all three complex fields $\cZ_A$ is given by:
\eqn\DimR{ \Delta_{\cZ}= R_{\cZ}={1\over 3} \,. }

As explained earlier, if one re-writes this \Neql2 theory and flow
in terms of \Neql1  superfields then some terms in the \Neql1
superpotential come directly from the \Neql2 superpotential whilst
others are related to the \Neql2 kinetic terms (D-terms).   The
former contain the relevant operators that drive the flow while
the latter, being related to kinetic terms of fields that are
frozen out,   become irrelevant in the IR and are simply dropped.

One can use this perspective in thinking about flows with two
non-zero and unequal  masses, $m_7\neq m_8$. There are various
classes of monomials in the superpotential \WNone\ before mass
terms are added: terms can be independent of $(\Phi^7,\Phi^8)$,
they can be linear or they can be quadratic in these fields:
\eqn\WSUThree{\eqalign{ &\widetilde{W}= g_{ m  n}  g_{ p q}
\eps^{abcd}\tcZbar^m_a \tcZ^n_b \tcZbar^p_c \tcZ^q_d \cr & +
\eps^{abcd}\Blp ( \tcZ^1_a \tcZ_b^2 \tcZ^3_c +\tcZbar^1_a
\tcZbar_b^2 \tcZbar^3_c ) \Phi_d^7+i( \tcZ^1_a \tcZ_b^2 \tcZ^3_c
-\tcZbar^1_a \tcZbar_b^2 \tcZbar^3_c ) \Phi_d^8\Brp \cr & +g_{ m
n}  \eps^{abcd} \tcZ^m_a \tcZbar^n_b \Phi^7_c \Phi^8_d. }}
where $m,n, \dots =1,2,3$,  and $ g_{ m  n}$ is some K\"ahler
metric.  The quadratic terms, coming from the final line in
\WSUThree, prevent one from integrating out both $\Phi^7$ and
$\Phi^8$ analytically.  However, it is precisely these quadratic
terms  in $(\Phi^7,\Phi^8)$ that have the form $\tcZ^4 \tcZbar^4$
and that become irrelevant in the \Neql2 flow. It therefore seems
reasonable to assume that these may be dropped in the general
family of flows.  Indeed, if we ignore these terms and  integrate
out $(\Phi^7,\Phi^8)$ in $\widetilde{W}+\Delta W_{BLG}$ we find:
\eqn\WSUThreeDef{\eqalign{ &\widehat{W}= g_{ m  n}  g_{ p q}
\eps^{abcd}\tcZbar^m_a \tcZ^n_b \tcZbar^p_c \tcZ^q_d \cr & + h_1
\blp \eps^{abcd} ( \tcZ^1_a \tcZ_b^2 \tcZ^3_c +\tcZbar^1_a
\tcZbar_b^2 \tcZbar^3_c )\brp ^2 +h_2 \blp \eps^{abcd}( \tcZ^1_a
\tcZ_b^2 \tcZ^3_c -\tcZbar^1_a \tcZbar_b^2 \tcZbar^3_c ) \brp^2 .
}}
This contains terms that are quartic as well as sextic and the two
parameters, $h_1$ and $h_2$, are the remnants of the mass
parameters.  We are unable to argue purely from the field theory
that this should flow to a SCFT in the IR, however, the gravity
dual suggests that it will flow to the $SU(3)\times U(1)$
symmetric \Neql2  point. This implies that in the IR $h_1=h_2$ and
that the quartic terms become tied by \Neql2\ supersymmetry to the
kinetic terms for the $\tcZ^m$ fields.

The main reason for not being able to provide an argument purely
from the field theory for the phase structure of this family of
flows is that in three dimensions, \Neql1 supersymmetry has no
${\cal R}$-symmetry and thus no chiral ring. However the main
feature of AdS/CFT is that the gravity dual can be used to study
strongly coupled field theory, and for the class of field theories
considered here we will see that the gravity dual provides much
sharper information about the phase structure.

%%%%%%%%%%%%%%%%%%%%%%%%%%%%%%%%%%%%%%%%
\newsec{Mass perturbations in maximal supergravity}
%%%%%%%%%%%%%%%%%%%%%%%%%%%%%%%%%%%%%%%%

%%%%%%%%%%%%%
\subsec{The scalars of gauged supergravity and their holographic
duals}
%%%%%%%%%%%%%

The $SU(3)$-invariant sector of gauged supergravity was studied
long ago in \refs{\WarnerVZ,\NicolaiHS,\WarnerDU}.  In terms of
the complex $4$-forms that parametrize the $E_{7(7)}/SU(8)$ of the
maximal theory, this six-dimensional sector may be parametrized as
follows.  Following \refs{\WarnerVZ,\WarnerDU}, introduce complex
coordinates, $(z_1, z_2 , z_3, z_4)$ on $\IR^8$ and define the
real forms:
\eqn\forms{\eqalign{ & J^{\pm}  ~\equiv~ {i\over 2}\,\Big(
\sum_{j=1}^3  dz_j \wedge d\bar z_j\Big) ~\pm~ {i\over 2}\, dz_4
\wedge d\bar z_4\,, \qquad  F_1^+  ~\equiv~  J^{+} \wedge J^{+}
\,,  \qquad  F_1^-  ~\equiv~  J^{-} \wedge J^{-}   \,,\cr &
F_2^{+} + i F_3^{+}    ~\equiv~  dz_1 \wedge d z_2   \wedge d z_3
\wedge d z_4 \,, \qquad F_2^{-} + i F_3^{-}  ~\equiv~  dz_1 \wedge
d z_2   \wedge d z_3  \wedge d \bar z_4 \,.}}
The forms $F_j^{+} $ and  $F_j^{-} $ are,  respectively, self-dual
and anti-self dual.  The $SO(8)$ of gauged supergravity acts on
$\IR^8$ as the vector representation and  there is  $SU(3)$
subgroup that leaves all these forms invariant.  There are also
two $U(1)$'s in $SO(8)$ that commute with this $SU(3)$ and rotate
the $z_j  \to e^{i \alpha} z_j$, $j =1,2,3$ and $z_4 \to e^{i
\beta} z_4$.  These $U(1)$ actions can be used to set
$F_3^{\pm}=0$.

These six four-forms may be viewed as defining six scalar fields
in ${\cal N}\!=\! 8$ supergravity and, as a sub-manifold of
${E_{7(7)} \over SU(8)}$,  they live in the coset
\eqn\Simpcoset{{SU(1 ,1)  \over U(1)} \times {SU(2,1) \over SU(2)
\times U(1)} \,,}
where $F_1^\pm$ defines the tangents to the first manifold and
$F_2^\pm$ and $F_3^\pm$ define the tangents on the second. We will
parametrize the scalar manifolds  using (complex) scalar fields
by, $w_j$, $j=1,2,3$,  with the $E_{7(7)}$ components given by:
\eqn\scalarparam{  \Sigma ~=~ \sum_{j=1}^ 3 \, \Big( {\rm Re}(w_j)
\,  F_j^{+} + i\, {\rm Im}(w_j) \,  F_j^{-}  \Big) \,,}
whose exponential form, in terms of the coset \Simpcoset,  reduces
to:
\eqn\cosetgauge{{\cal M}_1 ~=~ \exp \left( \matrix{ 0 & w_1 \cr
\bar w_1 & 0} \right) \,, \qquad {\cal M}_2~=~ \exp \left(
\matrix{ 0 & 0& w_2 \cr 0 & 0& w_3 \cr \bar w_2 & \bar w_3 & 0}
\right)  \,.}

The gauged supergravity theory in four dimensions contains $70$
scalar fields, and these are holographically dual to the
(traceless) bilinears in the scalars and fermions:
\eqn\bilinears{\eqalign{\cO^{IJ} ~=~ & \Tr~\big(X^I \, X^J) ~-~
\coeff{1}{8}\, \delta^{IJ}\, \Tr~\big(X^K \, X^K\big) \,, \quad
I,J, \dots =1,\dots,8\cr \cP^{AB} ~=~ & \Tr~\big(\lambda^A \,
\lambda^B\big) ~-~ \coeff{1}{8}\, \delta^{AB}\, \Tr~\big(\lambda^C
\, \lambda^C \big) \,, \quad A,B, \dots =1,\dots,8\,,}}
where $\cO^{IJ}$ transforms in the ${\bf 35}_s$  of $SO(8)$, and
$\cP^{AB}$ transforms in the ${\bf 35}_c$.    The real parts of
$w_j$ can be thought of as the duals  of $\cO^{IJ}$ for $I,J =7,8$
and the imaginary parts of $w_j$ can be thought of as the duals of
$\cP^{IJ}$ for $I,J =7,8$.  The real and imaginary parts of the
scalar, $w_1$, are separately invariant under distinct $SU(4)
\times U(1)$ groups, which means that $w_1$ is dual to the
following operator:
\eqn\wonehol{ w_1 \quad \leftrightarrow \quad  (\cO^{77} +
\cO^{88})   ~+~ i \, (\cP^{77} + \cP^{88})\,.  }
Similarly, $F_1^{+}+ i F_2^{+}$ is dual to ${\rm Tr} ((X^7 + i
X^8)^2)$ and  $F_1^{-}+ i F_2^{-}$
 is dual to ${\rm Tr} (( \lambda^7 + i  \lambda^8)^2)$. Thus
\eqn\otherwhol{ w_2 \quad \leftrightarrow \quad  (\cO^{77} -
\cO^{88})   ~+~ i\, (\cP^{77} - \cP^{88}) \,, \qquad w_3 \quad
\leftrightarrow \quad  \cO^{78}     ~+~ i\, \cP^{78} }
One can use the residual $U(1) \times U(1)$ invariance to
diagonalize the  fermion and boson mass matrices and take $w_3 =
0$. To get the $G_2$ invariant critical point and flows one takes
$w_1 = \pm  w_2, w_3 = 0$, while for the $SU(3)$ invariant
critical point and flow one takes ${\rm Im}(w_1)= {\rm Re}(w_2) =
w_3 = 0$ \WarnerVZ.

%%%%%%%%%%%%%
\subsec{The scalar action of gauged supergravity}
%%%%%%%%%%%%%

To exponentiate the scalar matrices, it is convenient to use a
polar parametrization and take (for $w_3 =0$):
\eqn\polars{ w_1 ~=~ \lambda \, e^{-2 i \phi}  \,, \qquad   w_2
~=~ \coeff{1}{2} \,\chi  \,  e^{i \varphi}    \,. }
After exponentiating one can write the matrices \cosetgauge\ in
terms of the scalar fields: 
\eqn\polars{ \zeta_1 ~=~ \tanh  \lambda \, e^{-2 i \phi}  \,,
\qquad   \zeta_2 ~=~ \tanh (\half \chi) \,  e^{i \varphi}    \,,}
for which the supergravity Lagrangian \deWitIG\ gives the kinetic
term:
\eqn\cosetkin{ \cL_{kin.}~=~ -  \bigg[3\, { \nabla_\mu \zeta_1 \,
\nabla^\mu \bar \zeta_1  \over (1 - |\zeta_1|^2)^2 } ~+~ 4\,
\sum_{j =2}^3 { \nabla_\mu \zeta_j \, \nabla^\mu \bar \zeta_j
\over (1 -( |\zeta_2|^2 + |\zeta_3|^3))^2) } \bigg] \,,}
where we have restored $\zeta_3$ via symmetry.  In terms of the
polar representation one has:
\eqn\polarkin{\eqalign{
 \cL_{kin.}& ~=~ -K_{i j} \,(\nabla_\mu \psi^i) (\nabla^\mu  \psi^j)  \cr
& ~=~ -  \big[   ( \partial_\mu \chi )^2  ~+~  \sinh^2 \chi \,  (
\partial_\mu \varphi)^2    ~+~
 3 \, \big( ( \partial_\mu \lambda )^2  ~+~   \sinh^2 2\lambda\,  (\partial_\mu \phi)^2  \big) \big] \,,}}
where  $K_{ij}$ is the metric on the scalar space with $\psi^i  =(
\lambda, \chi, \phi, \varphi)$.

Following \refs{\FreedmanGP\WarnerKH\AhnAQ{--}\KhavaevGB}, a
superpotential can be extracted from the eigenvalues of the
$A_1$-tensor that appears in the variation of the  gravitino of
the $\cN\!=\!8$ theory \deWitIG.  In the $SU(3)$ invariant sector
there are two candidate eigenvalues\foot{The $SU(3) \times U(1)$
invariant sector is given by taking  $\zeta_1$ to be real and $\zeta_2$ to be purely
imaginary, and hence these two eigenvalues  are equal.}  \AhnMF \
that are related by  $\zeta_2 \to - \bar \zeta_2$.   Choosing one
of these eigenvalues, we define the complex superpotential, ${\cal
W}$, by
\eqn\superpot{{\cal W}~=~   (1 - |\zeta_1|^2  )^{-{3\over 2}}(1 -
|\zeta_2|^2)^{-2}  \Big[(1 + \zeta_1^3)(1 + \zeta_2^4)  ~+~ 6\,
\zeta_1 \zeta_2^2 (1 + \zeta_1)  \Big] \,.}

The supergravity potential on the $SU(3)$ invariant sector
\WarnerVZ\ is then given by \AhnMF:
\eqn\sugrpotcplx{ \eqalign{{\cal P}~=~&  2\, g^2 \bigg[
\Big|{\partial \cW \over \partial \chi}\Big|^2  +  {4 \over 3}
\Big|{\partial \cW \over \partial \lambda} \Big|^2 -  3\,  |\cW|^2
\bigg]  \cr   ~=~&  2\, g^2 \bigg[ {4 \over 3} (1 - |\zeta_1|^2)^2
\Big|{\partial \cW \over \partial \zeta_1}\Big|^2  + (1 -
|\zeta_2|^2)^2  \Big|{\partial \cW \over \partial \zeta_2} \Big|^2
-  3\,  |\cW|^2 \bigg]  \,.} }
The real superpotential is given by $|\cW|$ and one also has:
\eqn\sugrpot{ {\cal P}~=~  2\, g^2 \bigg[ \Big({\partial |\cW|
\over \partial \chi} \Big)^2  + {1 \over \sinh^2 \chi } \,
\Big({\partial |\cW| \over \partial \varphi} \Big)^2 +  {1\over 3}
\Big({\partial |\cW| \over \partial \lambda} \Big)^2 +  {1\over 3
\,   \sinh^2 2\lambda} \,  \Big({\partial |\cW| \over \partial
\phi} \Big)^2 - 3\,  |\cW|^2 \bigg]   \,.}
This is a consequence of identities that come from the fact that
$\cW$ is holomorphic up to an overall pre-factor:
\eqn\cWids{ {\partial_\phi} \log  \cW - i \sinh 2\lambda \,
{\partial_ \lambda} \log  \Big( { \cW \over |\cW|} \Big)~=~  0 \,,
\quad  {\partial_\varphi} \log  \cW + i \sinh \chi \,
{\partial_\chi} \log  \Big( { \cW \over |\cW|} \Big)~=~  0\,.}

The superpotential has an $SO(8)$-invariant critical point, with
$\cN\!=\!8$ supersymmetry, at $\zeta_1 = \zeta_2 =0$ and with
cosmological constant, $ \Lambda = -6 g^2$.  The $SU(3)\times
U(1)$-invariant critical point, with $\cN\!=\!2$ supersymmetry is
given by:
\eqn\SUpoint{\eqalign{ &\lambda = \lambda_2 \equiv \coeff{1}{4}
\log(3) \,, \quad    \chi = \pm\chi_2 \equiv \pm \log\Big(
{\sqrt{3}-1 \over \sqrt{2}}\Big) \,, \quad   \phi = \phi_2 \equiv
0\,, \quad    \varphi = \pm \varphi_2 \equiv \pm {\pi \over 2}\,;
\cr
& \Lambda_{SU(3)} ~=~  -{9\sqrt{3}  \over 2}\,g^2 ~\approx~  -  7.79423 \, g^2\,,}}
with all possible choices of signs.    In terms of the complex
variables this corresponds to:
\eqn\SUpointcplx{ \zeta_1 ~=~  2 -  \sqrt{3}   \,, \qquad
\zeta_2~=~ \pm i ( \sqrt{3}  - \sqrt{2})   \,.}
 The $G_2$-invariant critical
point, with $\cN\!=\!1$ supersymmetry is given by:
\eqn\Gtwopoint{ \eqalign{ & \lambda =  \pm \coeff{1}{2} \chi =
\coeff{1}{2}  \chi_1 \equiv  \coeff{1}{4} \log
\Big(\coeff{1}{5}\Big( 1 + 4\sqrt{3} + 2 \sqrt{2} \sqrt{3
+\sqrt{3}} \,  \Big) \Big)\,, \cr & \phi =    -\coeff{1}{2}
\varphi = \phi_1 \equiv -\coeff{1}{2}  \varphi_1 \equiv
\coeff{1}{2}  \arccos \bigg( {\sqrt{3 - \sqrt{3}} \over 2} \bigg)
\,;  \cr &  \Lambda ~=~ -{216 \, \sqrt{2}\, 3^{1 \over 4}   \over
25 \, \sqrt{5}}\,g^2 ~\approx~ -7.19158 \,g^2 \,.}}
There is also a solution with $\chi \to \chi + \pi$.  The $G_2$
critical points are given by $\zeta_2  = \pm \zeta_1^{\pm 1} $ for
all choices of sign.  The actual values of $\zeta_1$ are a rather
unedifying mess.

%%%%%%%%%%%%%
\subsec{The supersymmetric flow equations}
%%%%%%%%%%%%%

To set up a supersymmetric flow one takes the four-dimensional
metric to have the form:
\eqn\fourmet{ds_{1,3}^2 ~=~ dr^2 ~+~ e^{2 A(r)}\big( \, \eta_{\mu
\nu} \, dx^\mu \, d x^\nu \big) \,.}
We take the Lagrangian of the scalars coupled to gravity to be:
\eqn\Lag{  \cL~=~  \half R ~-~  \cP ~+~ \cL_{kin.} \,.}
The supersymmetric flow equations are then obtained from the
supersymmetry variations of the fermions and one finds \AhnMF:
\eqn\floweqs{ \eqalign{ {d \lambda \over d r} ~=~  & \pm{
\sqrt{2}\, g \over 3} \,  \partial_\lambda |\cW|  \,, \qquad {d
\chi \over d r} ~=~    \pm \sqrt{2} \,g \,  \partial_\chi  |\cW|
\,,  \cr {d \phi \over d r} ~=~  & \pm { \sqrt{2}\, g\over 3
\sinh^2 2\lambda}  \,  \partial_\phi  |\cW|   \,, \qquad {d
\varphi \over d r} ~=~   \pm { \sqrt{2}\, g\over  \sinh^2  \chi}
\, \,  \partial_ \varphi  |\cW|   \,,   \qquad    {d A \over d r}
~=~  \mp \sqrt{2} \,g \, |\cW| \ .}}
The equations for the flow of the scalars may be rewritten in
terms of the scalar metric:
\eqn\floweqsmet{ {d \psi^i  \over d r} ~=~  \pm \sqrt{2}\, g  \,
K^{ij} \, {\partial |\cW | \over \partial  \psi^j}  \,,   }
where $K^{ij}$ is the inverse of the metric $K_{ij}$ defined in
\polarkin.  In terms of the complex coordinates \floweqs\ become
\eqn\floweqscplx{ {d \zeta_1 \over d r} ~=~  \pm{ 2\sqrt{2}\, g
\over 3} \,  (1 - |\zeta_1|^2 )^2 \, {\partial \cW \over \partial
\bar  \zeta_1}  \,, \qquad{d \zeta_2 \over d r} ~=~   \pm{  g
\over \sqrt{2}} \,  (1 - |\zeta_2|^2 )^2 \, {\partial \cW \over
\partial \bar  \zeta_2} \,.  }

Given the cosmological constants of the three supersymmetric
critical points, they suggest  a possible flow, by steepest
descent from the $G_2$ point to the $SU(3) \times U(1)$ point. One
can see graphically that this is  possible.  The superpotential,
$|\cW|$, depends upon four variables and the easiest way to see
the critical points is to create a function of two variables,
$|\widehat \cW(\chi,\lambda)|$  by substituting the following into
$\cW$:
\eqn\interp{
 \phi =  \phi_2 + (\phi_1-\phi_2)   { (\chi^2 - \chi_2^2)  \over (\chi_1^2 - \chi_2^2) } \,, 
 \qquad  \varphi =  \varphi_2 + (\varphi_1-\varphi_2)   { (\chi^2 - \chi_2^2)  \over (\chi_1^2 - \chi_2^2) } \,,  }
where the $\chi_j$, $\phi_j$ and $\varphi_j$ are defined in
\Gtwopoint\ and \SUpoint.  This substitution ensures that the
function $|\widehat \cW|$ slices through the critical points of $|
\cW|$.  The result is depicted in Fig. 2.

%%%%%%%%%%%%%%%%%%%%%%%%%%%%%%%
\goodbreak\midinsert
\vskip .2cm \centerline{ {\epsfxsize 5cm \epsfbox{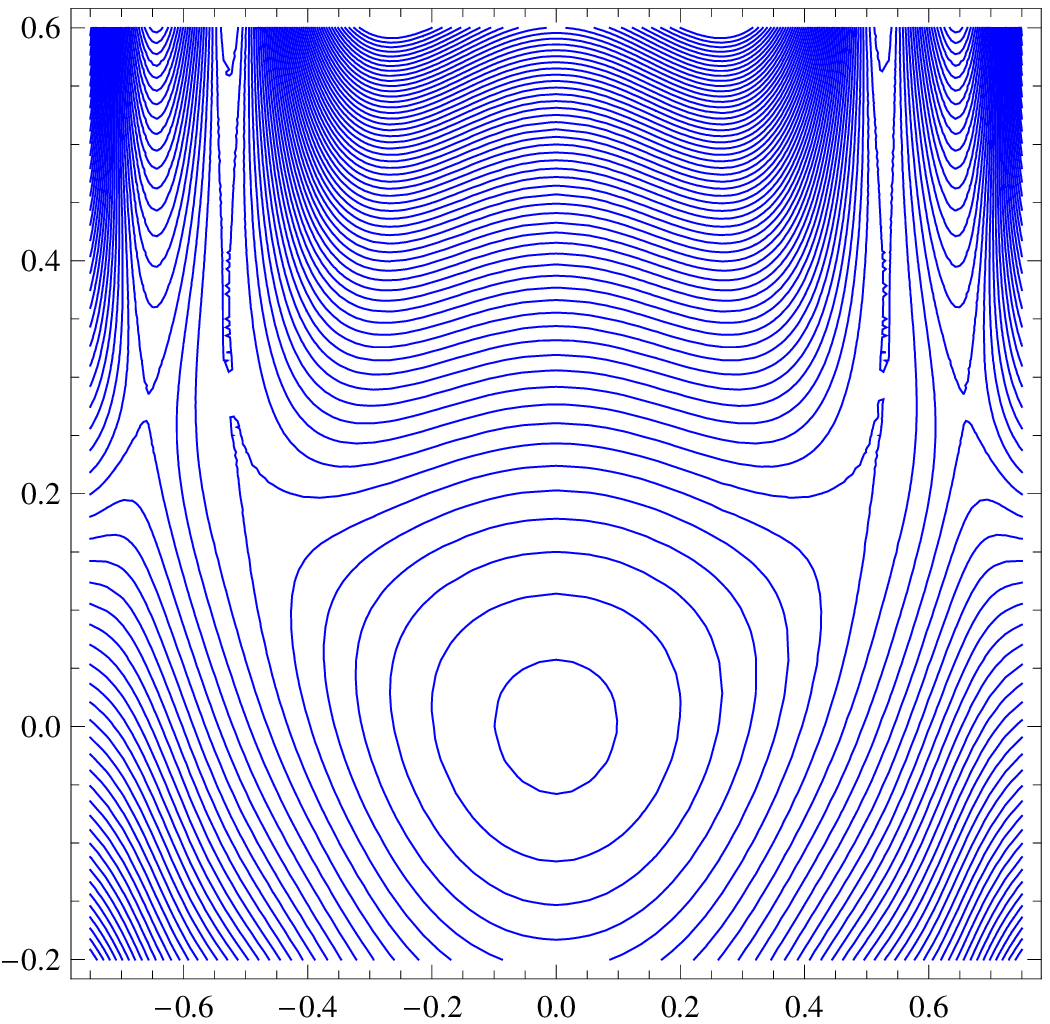}}
\hskip0.5cm  {\epsfxsize 6cm\epsfbox{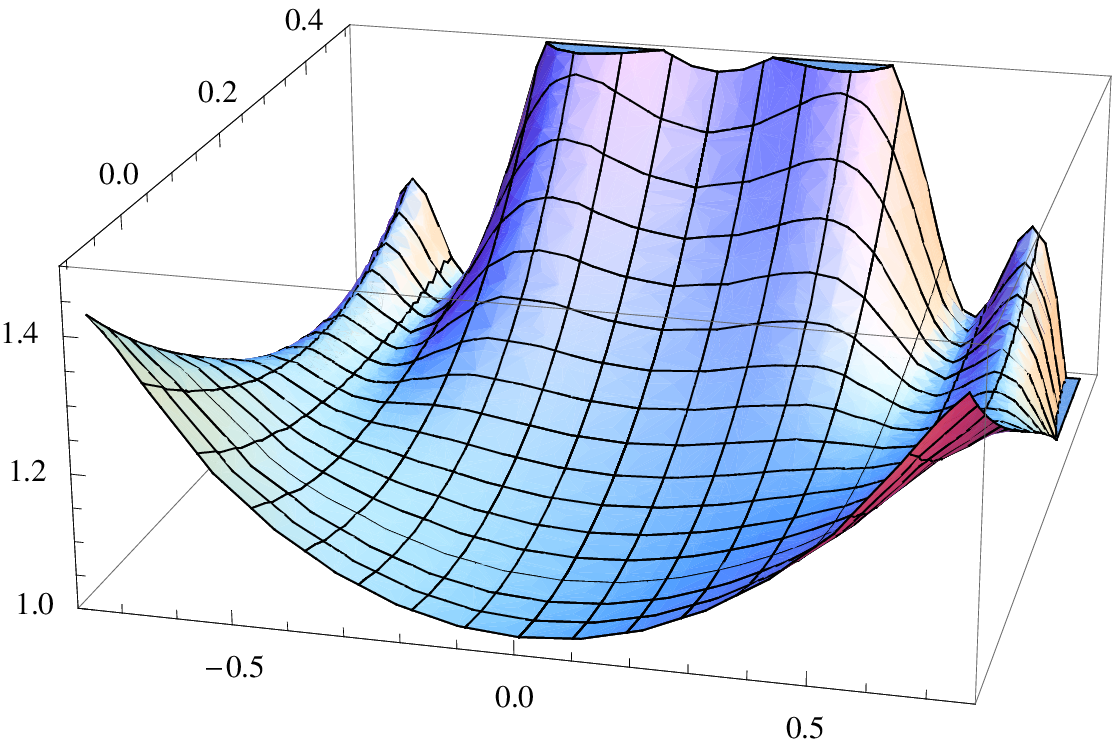}}} \vskip 0.2cm
\leftskip 2pc
 \rightskip 2pc\noindent{\ninepoint\sl
{\bf Fig.~2}: \baselineskip= 8pt   Plots of the function
$|\widehat \cW(\chi,\lambda)|$ obtained by making the
substitutions \interp\ into the superpotential $|\cW|$.  The
left-right axis is $\chi$ and the other axis is $\lambda$. There
are five critical points visible and they are related by $\chi \to
-\chi$. The SO(8) invariant critical point is the central minimum.
Moving away from this, the first saddle points are the
$G_2$-invariant critical points and the second pair of highest
saddles  are the $SU(3) \times U(1)$-invariant critical points.}
\endinsert
%%%%%%%%%%%%%%%%%%%%%%%%%%%%%%%

There is a unique steepest  descent on the superpotential $| \cW|$
that goes from the $SU(3) \times U(1)$-invariant critical point to
the $G_2$ invariant critical point.  One can also find a family of
steepest descent flows  on  $| \cW|$ starting from the $SU(3)
\times U(1)$-invariant critical point and descending ultimately to
the $SO(8)$-invariant critical point.   There is the direct
descent, which preserves $SU(3) \times U(1)$, and there are
descents that approach the $G_2$ invariant fixed point first
before turning down to the $SO(8)$-invariant critical point.
Indeed one may approach  $G_2$ invariant fixed point  arbitrarily
closely.  These are depicted in Fig. 3.

 %%%%%%%%%%%%%%%%%%%%%%%%%%%%%%%
\goodbreak\midinsert
\vskip .2cm \centerline{ {\epsfxsize 3in\epsfbox{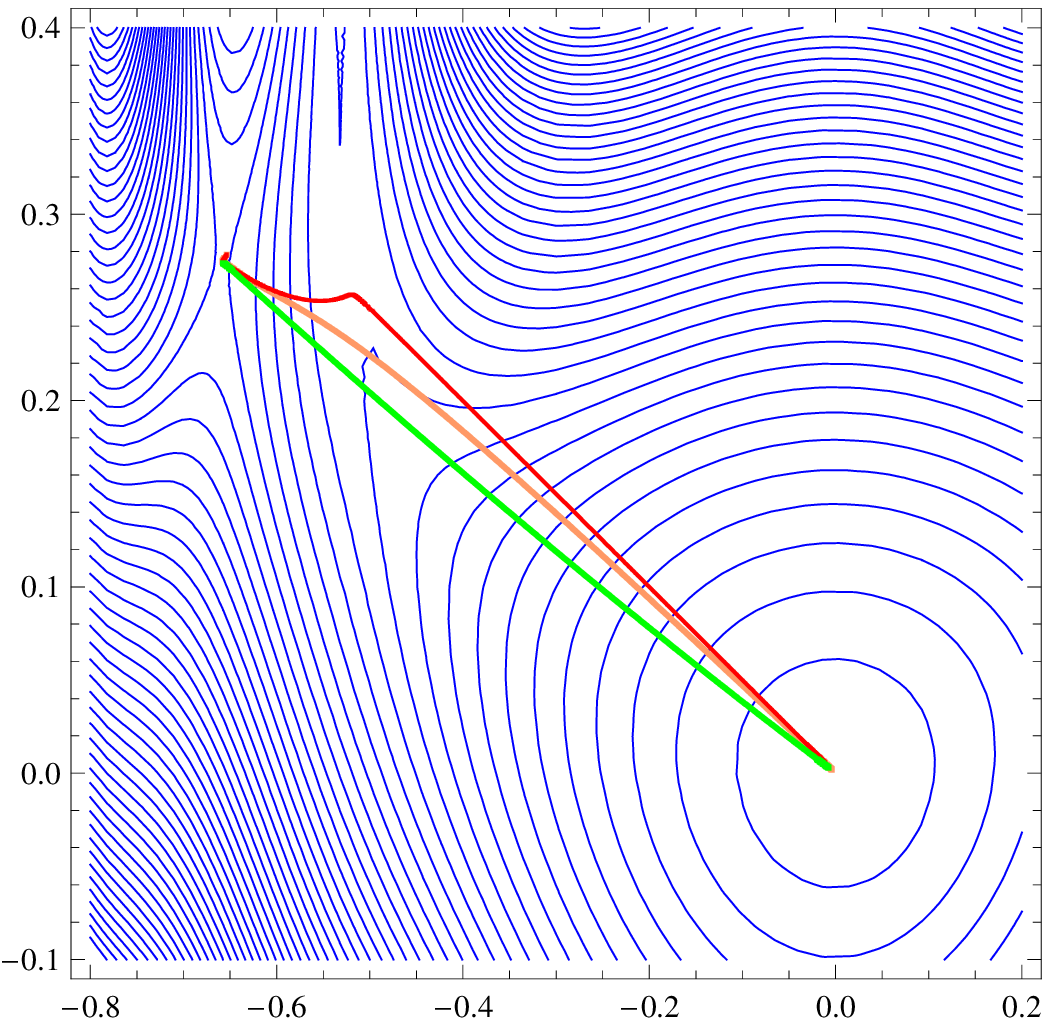}}}
\vskip 0.2cm \leftskip 2pc \rightskip 2pc\noindent{\ninepoint\sl
\baselineskip=8pt {\bf Fig.~3}:   This shows details of the
contour plot of $|\widehat \cW|$ in Fig. 2.  Three steepest
descent paths shown:  One going directly from the $SU(3) \times
U(1)$-invariant critical point to the $SO(8)$-invariant critical
point.  Another goes from the $SU(3) \times U(1)$-invariant
critical point and passes extremely close to the $G_2$-invariant
critical point before descending to the $SO(8)$-invariant critical
point.  The third is a generic intermediate path between these
extremes. The physical holographic RG flows follow these
trajectories in reverse.  Note also that there is some relative
distortion of the paths and the contours because the paths
represent numerical solutions on the complete superpotential,
$|\cW|$, while the contours are those of $|\widehat \cW|$.}
\endinsert
%%%%%%%%%%%%%%%%%%%%%%%%%%%%%%%

The field theory flows are, of course, steepest descents on
$-|\cW|$ and therefore flow in the opposite direction to the
foregoing discussion.  The $G_2$ flow corresponds to tuning $m_1
\ne 0, m_2=0$, while the $SU(3) \times U(1)$ invariant flow
corresponds to  $m_1= m_2$.  From the supergravity it is evident
that if one has the $G_2$ flow with $m_1 \ne 0$ and if one turns
on a small value for $m_2$, then the flow is deflected to the
$SU(3) \times U(1)$ invariant fixed point and so $m_2$ grows until
$m_2 =m_1$. Generic flows out of all of the fixed points typically
run off to infinity, or ``Hades,''  and this simply means that the
Coulomb branch is dominating the infra-red end of the flow
\refs{\FreedmanGP, \FreedmanGK}.  The interesting new feature is
that there is a ``cone,'' or family, of flows bounded by the
$SU(3) \times U(1)$ and  $G_2$ invariant flows and whose infra-red
limit is the $SU(3) \times U(1)$ invariant fixed point.

%%%%%%%%%%%%%
\subsec{Flows near the critical points}
%%%%%%%%%%%%%

To understand the pattern of the flows around the three fixed
points, it is instructive to compute the scaling dimensions of the
operators in the $SU(3)$-invariant sectors and see how they govern
the flows.  This requires the linearization of the flow equations
in the neighborhood of the fixed points.

For the $SO(8)$-invariant fixed point, the polar coordinate system
is singular and it is more convenient to linearize \floweqscplx\
which, around $\zeta_j =0$, give:
\eqn\flowlinSO{ {d \zeta_1 \over d r} ~\approx~  \pm \sqrt{2}\,
g\,  \zeta_1 + \dots  \,, \qquad  {d \zeta_2 \over d r}~\approx~
\pm \sqrt{2}\, g\,  \zeta_2 + \dots  \,,   \qquad    {d A \over d
r} ~\approx~  \mp \sqrt{2} \,g+ \dots     \,.  }
The canonical form of the AdS metric of radius $L$ is to take:
\eqn\AdSA{  A(r) ~=~ e^{r/L}   \,,}
which means that modes are non-normalizable if they behave as:
\eqn\AdSscaling{ e^{-\Delta{r / L }}\,, \quad \Delta \le {3 \over
2}  \,.}
For the flows \flowlinSO\ one has:
\eqn\rndscales{  g  ~=~ \mp  {1 \over\sqrt{2}\, L}   \,, \qquad
\zeta_1 ~=~ a_1 \, e^{-{r / L}} \,, \qquad \zeta_2 ~=~ a_2 \,
e^{-{r/ L}}  \,.}
for some constants $a_j$, and so these modes are all
non-normalizable. Thus they represent mass insertions into the
Lagrangian and not vevs of background fields.   There is an
ambiguity in the holographic dictionary if a field that has
dimension $\Delta$ has a supergravity mode that scales as:
\eqn\scaledic{  e^{-\Delta{r / L }}\, \quad {\rm or} \quad e^{-(3-
\Delta){r / L }} \,.}
For the fermionic and bosonic mass terms one has $\Delta =2$ and
$\Delta =1$ and these correspond to \rndscales\ provided that the
fermions and bosons correspond to different choices in \scaledic.

One expects terms in the Lagrangian that are related by
$\cN\!=\!1$ supersymmetry to have scaling dimensions that differ
by $1$.   One should note that, when this is translated through
the foregoing holographic dictionary, the dual supergravity
scalars in a supermultiplet are expected to have exponents that
either differ by $1$ or that sum to $2$.   The exponents in
\rndscales\ have the latter behavior.

In the neighborhood of the other two fixed points it is convenient
to use the polar form of the Lagrangian and linearize \floweqsmet.
Indeed, one obtains the canonical AdS metric, \AdSA\ if one now
takes
\eqn\gLreln{  g  ~=~ \mp  {1 \over\sqrt{2}\, L_* \, |\cW|_*} \,,}
where $|\cW|_*$ is the value of the superpotential  at the
critical point and $L_*$ is the $AdS$ radius corresponding to the
fixed point.  The linearization of  \floweqsmet\ is then
\eqn\linfloweqs{ {d \psi^i  \over d r} ~=~  -  {1 \over  L_*}
\,{\cM^i}_k\, (\psi^k - \psi_0^k) \,,   \qquad  {\cM^i}_k~\equiv~
\bigg( {K^{ij} \over  |\cW| }\,{\partial^2 |\cW | \over \partial
\psi^j\,  \partial  \psi^k} \bigg)_*\,,}
where $*$ denotes the value at the critical point.  Therefore, we
need the eigenvalues of the matrix $ {\cM^i}_k$.

At the $SU(3)\times U(1)$ invariant fixed point the eigenvalues of
$ {\cM^i}_k$ are:
\eqn\SUevs{ \eqalign{ \big(\coeff{1}{2} \big(1 -\sqrt{17}\big) ,
\coeff{1}{2} \big(3 -\sqrt{17}\big),&  \coeff{1}{2}
\big(\sqrt{17}+ 1  \big) ,  \coeff{1}{2} \big(\sqrt{17}+3 \big)
\big) \cr &~\approx~ (-1.56155, -0.561553,   2.56155,  3.56155)
\,,} }
and at the $G_2$ invariant fixed point the eigenvalues of $
{\cM^i}_k$ are:
\eqn\Gtwoevs{ \eqalign{ \big( 1 - \sqrt{6} , 1-\coeff{1}{\sqrt{6}}
,  &  \big(\coeff{1}{\sqrt{6}} + 1  \big) , \big(\sqrt{6}+ 1 \big)
\big) \cr &~\approx~ (-1.44949,   0.591752, 1.40825,  3.44949)
\,.} }
Note that at each point the eigenvalues come in pairs that add to
$2$, consistent with $\cN\!=\!1$ supersymmetry. Negative
eigenvalues correspond to irrelevant operators that flow into the
fixed point in the infra-red.  There is one such operator for the
$G_2$-invariant point, corresponding to the flow from the $SO(8)$
invariant fixed point.   There are two such operators for the
$SU(3)\times U(1)$ point and these correspond to the family, or
cone,  of flows that arrive at the $SU(3)\times U(1)$ point from
the $SO(8)$ point.  The positive eigenvalues correspond to
relevant operators or to vevs that drive the flow away from the
fixed point in the infra-red.  The  two positive eigenvalues at
the $SU(3)\times U(1)$ point are greater than ${3 \over 2}$, which
means the modes are normalizable and therefore correspond to
perturbations of the state of the system.  Based upon the
experience of \refs{\FreedmanGP,\KhavaevGB,\FreedmanGK}, it seems
reasonable to expect that they correspond to some form of Coulomb
branch flow.

At the $G_2$ point there are three positive eigenvalues.  One of
them is normalizable and presumably corresponds to a Coulomb
branch flow but, in contrast to the analogous situation in
four-dimensional Yang-Mills theory, the scaling dimension is
greater than $3$ and so this ``Coulomb flow'' is being driven by
the vev of an irrelevant operator. More interesting are the two
other eigenvalues, $1 \pm {1 \over \sqrt{6}}$, which correspond to
non-normalizable modes, and hence must represent perturbations of
the Lagrangian. Note that these eigenvalues sum to $2$ and thus
may be interpreted as supersymmetric counterparts of one another.
Indeed, they must represent the fermionic and bosonic mass terms
that generate the $\cN\!=\!1$ supersymmetric flow from the $G_2$
point to the $SU(3)\times U(1)$ point. Supergravity therefore
predicts the dimensions of the corresponding operator to be  $1
\pm  {1 \over \sqrt{6}}$ and $2 \pm {1 \over \sqrt{6}}$, and hence
there is an anomalous dimension of  $  \pm {1 \over \sqrt{6}}$. It
is interesting that the dimensions are not rational, but this is
entirely possible since there is no continuous $\cR$-symmetry to
protect operator dimensions.  It would be most interesting to see
if there is a way to compute these relevant operator dimensions
directly within the field theory.

%%%%%%%%%%%%%%%%%%%%%%%%%%%%%%%%%%%%%%%%
\newsec{Final comments}
%%%%%%%%%%%%%%%%%%%%%%%%%%%%%%%%%%%%%%%%

We have studied the field theory on a stack of membranes  by deforming the 
theory with mass terms. The specific mass terms we considered trigger flows that terminate at
superconformal Chern-Simons matter theories in the IR. The phase
structure of the general flow in this class is hard to study
directly in the field theory since only \Neql1 supersymmetry is
preserved. Nevertheless our study of the gravity dual provides a
compelling description of these flows.

The remaining challenges directly related to this family of flows
lie in the field theory. For instance, simply calculating the
dimension of operators at the $G_2$ symmetric point and comparing
them to the supergravity spectrum would be an important
achievement since there is no holomorphy in the field theory and
it is strongly coupled.

There are many other \Neql1 supersymmetric mass terms
that can be considered and it would be interesting to study these
using holography. One particular class of these flows involves an
equal mass term for all four complex scalars and was considered
from the gravity point of view \refs{\BenaZB, \PopeJP, \BenaJW} and  
from the field theory point of view \refs{\LinNB,\GomisCV}. A
related, non-holomorphic mass deformation was studied in
\refs{\GomisVC}. This flow preserves sixteen supercharges and has a
number of isolated vacua; it remains unsolved how to count these
vacua correctly from the field theory. The difficulty in studying
these mass deformations in the ABJM model is the same difficulty
we have encountered in the current work, namely that the mass
terms preserve the supersymmetry which is not manifest in the ABJM
model.

There are also flows with equal mass terms for two complex scalars and  preserving 
eight supersymmetries.  These have been studied in  \refs{\GowdigereUK, \GowdigereJF}
and can be considered as the analogue of the \Neql{2^*}  mass
deformation of \Neql4 SYM in four dimensions \refs{\PilchUE}, which
flows to large-$N$ Donagi-Witten theory \refs{\DonagiCF}. 
Another family of flows that has not been examined closely in supergravity is the  
deformation with equal masses  for three complex scalars. This should also preserve
$SU(3)\times U(1)$ symmetry but should not terminate at an SCFT.  This
high level of symmetry should also be sufficient to make it amenable to study from
the supergravity perspective, perhaps even calculating the full
eleven-dimensional solution. Moreover, the corresponding field theory should
be related to the compactification to three dimensions of \Neql{1^*} Yang-Mills theory, obtained 
by giving masses to the three chiral multiplets in  \Neql4 Yang-Mills theory.  The corresponding
field theory in $(2+1)$ dimensions has been extensively studied and used to compute
exact elliptic superpotentials  \refs{\DoreySJ, \DoreyFC}.  Given the new developments in the field 
theory on the M2 branes it would   be very interesting to revisit these earlier results and see how the
are related via massive flows..  

It would also be very interesting to uplift the RG flow solutions that
we found in four-dimensional gauged supergravity to eleven
dimensions. This has been already done for the ${\cN}=2$ flow
which corresponds to $m_1=m_2$ in \refs{\CorradoNV}. It is well
known how to uplift the metric of solutions to four-dimensional
$\cN=8$ gauged supergavity to eleven dimensions \refs{\deWitNZ},
however the techniques for finding the internal
fluxes are rather cumbersome \deWitIY. One of the non-trivial features of the solution in
\refs{\CorradoNV} is the presence of internal four-form flux and
one can expect that such flux will be present for the whole
$SU(3)$ invariant family of flow solutions discussed here. The
solutions with $m_1\neq m_2$ will have also smaller internal
symmetry group and less supersymmetry which makes the
eleven-dimensional uplift a non-trivial task.

In terms of string compactifications, $AdS_4$ vacua are
phenomenologically interesting for many reasons. It would be
interesting to develop a better understanding of such backgrounds
which preserve only two supercharges and the dual three-dimensional
field theory is presumably a useful place to perform such studies.
As such, \Neql1 CS-matter theories, like the ones studied in
this paper, may be an appropriate place to start.

\bigskip
%%%%%%%%%%%%%%%%%%%%%%%%%%%%%%%%%%
\leftline{\bf Acknowledgments}
%%%%%%%%%%%%%%%%%%%%%%%%%%%%%%%%%%

NH would like to thank Michela Petrini for discussions. The work
of NB, KP and NPW is supported in part by the DOE grant
DE-FG03-84ER-40168. The work of NH is supported by DSM CEA-Saclay
and the grant Marie Curie IRG 046430. NB is supported also by a
Graduate Fellowship from KITP and in part by the National Science
Foundation under Grant No. PHY05-51164.

%%%%%%%%%%%%%%%%%%%%%%%%%%%%%%%%%%
\appendix{A}{Projecting  \Neql2 to  \Neql1 superspace in three dimensions}
%%%%%%%%%%%%%%%%%%%%%%%%%%%%%%%%%%

Here we summarize some aspects of how to break up the
three-dimensional \Neql2  superfields into \Neql1 superfields, a
complete description is given in \refs{\AvdeevJT,\AvdeevZA}.  This
is useful as it allows one to consider an \Neql2 action and add
\Neql1 preserving operators to it.

The complex spinor of \Neql2   superspace is decomposed as\foot{The
irreducible spinor $\tha^\al$ in three dimensions has two real components but
often its complex counterpart with four real components is also
denoted $\tha^\al$. We hope this will not cause too much
confusion.}  $\tha = \tha_1+i \tha_2$ and  to reduce the action to
\Neql1 supersymmetry we integrate out the $\tha_2$ dependance.
First we decompose the $N=2$ differentials in terms of \Neql1
differentials
\eqn\Dtwoone{ D_\al = \half(D_{1\al}+i D_{2\al}),\ \ \Dbar^\al =
\half(D^\al_{1}-i D^\al_{2} ). }
This allows us to write the superspace \Neql2 measures in a way
which then facilitates the reduction of the action to \Neql1
superspace
\eqn\measure{\eqalign{ \int d^3x d^4\tha =&-\int d^3x D_1^2 D_2^2,
\cr \int d^3x d^2\tha =& \int d^3x D_1^2. }}
Then the \Neql2 fields reduce to \Neql1 fields as
\eqn\Ntwofields{\eqalign{ \cZ |_{\tha_2=0} = \tcZ, &\ \ \ \cZbar
|_{\tha_2=0} = \tcZbar, \cr \cV|_{\tha_2=0}=0&\ \ \ D_{2\al} \cV
|_{\tha_2=0} = \Gam_\al,\ \ \ D_2^2 \cV|_{\tha_2=0}=R }}
where $\tcZ$ is a {\it complex} \Neql1  scalar superfield and $R$
is {\it real} \Neql1  scalar superfield. Since \Neql1  superspace
is real, we can break a complex scalar superfield into real and
imaginary parts and in section 2 we used the complex structure
\eqn\Zsagain{\eqalign{ \tcZ^1= \Phi^1+i \Phi ^2, &\ \ \  \tcZ^2=
\Phi ^3+i \Phi ^4 \,, \cr \tcZ^3= \Phi ^5+i \Phi ^6, &\ \ \
\tcZ^4= \Phi ^7+i \Phi ^8 \, }}
where $\Phi^i$ are real \Neql1 superfields.

In three dimensions, the \Neql2 gauge superfield $\cV$ is a
bosonic superfield while the \Neql1 gauge superfield  $\Gam^\al$
is a fermi superfield. In the CS matter theories studied in this
paper, integrating out the auxiliary superfield $R$ will result in
additional \Neql1 superpotential terms.

\listrefs
\end